\pdfoutput=1 

\documentclass[conference]{IEEEtran}
\IEEEoverridecommandlockouts
\usepackage{cite}
\usepackage{amsmath,amssymb,amsfonts}
\usepackage{graphicx}
\usepackage{textcomp}
\usepackage{xcolor}
\usepackage{amsthm,amssymb}
\usepackage{adjustbox}
\usepackage{algorithm}
\usepackage{algorithmic}
\usepackage{bm}
\usepackage{booktabs}
\usepackage{url}
\usepackage{multirow}
\usepackage{array}
\usepackage{float}
\usepackage{atbegshi,picture}
\usepackage{lipsum}
\usepackage[bottom]{footmisc}

\usepackage{tikz,pgfplots}
\usetikzlibrary{pgfplots.groupplots}
\usetikzlibrary{arrows.meta}
\usetikzlibrary{patterns}
\usetikzlibrary{matrix}

\newcommand*\circled[1]{\tikz[baseline=(char.base)]{
            \node[shape=circle,fill,inner sep=0.8pt] (char) {\textcolor{white}{#1}};}}
\newcommand{\STAB}[1]{\begin{tabular}{@{}c@{}}#1\end{tabular}}

\begin{document}
\title{Model-Architecture Co-Design for High Performance Temporal GNN Inference on FPGA}

\author{
\IEEEauthorblockN{Hongkuan Zhou\IEEEauthorrefmark{1}\textsuperscript{\textsection},Bingyi Zhang\IEEEauthorrefmark{1}\textsuperscript{\textsection},Rajgopal Kannan\IEEEauthorrefmark{2},Viktor Prasanna\IEEEauthorrefmark{1},Carl Busart\IEEEauthorrefmark{2}}
\IEEEauthorblockA{
    \IEEEauthorrefmark{1}University of Southern California \IEEEauthorrefmark{2}US Army Research Lab\\
    \IEEEauthorrefmark{1}\{hongkuaz,bingyizh,prasanna\}@usc.edu \IEEEauthorrefmark{2}\{rajgopal.kannan.civ,carl.e.busart.civ\}@army.mil}
}
\maketitle
\begingroup\renewcommand\thefootnote{\textsection}
\footnotetext{Equal contribution}
\endgroup

\begin{abstract}
Temporal Graph Neural Networks (TGNNs) are powerful models to capture temporal, structural, and contextual information on temporal graphs. The generated temporal node embeddings outperform other methods in many downstream tasks. Real-world applications require high performance inference on real-time streaming dynamic graphs. However, these models usually rely on complex attention mechanisms to capture relationships between temporal neighbors.
In addition, maintaining vertex memory suffers from intrinsic temporal data dependency that hinders task-level parallelism, making it inefficient on general-purpose processors. 
In this work, we present a novel model-architecture co-design for inference in memory-based TGNNs on FPGAs.
The key modeling optimizations we propose include a light-weight method to compute attention scores and a related temporal neighbor pruning strategy to further reduce computation and memory accesses. These are holistically coupled with key hardware optimizations that leverage FPGA hardware. We replace the temporal sampler with an on-chip FIFO based hardware sampler and the time encoder with a look-up-table.
We train our simplified models using knowledge distillation to ensure similar accuracy vis-\'a-vis the original model. Taking advantage of the model optimizations, we propose a principled hardware architecture using batching, pipelining, and prefetching techniques to further improve the performance. We also propose a hardware mechanism to ensure the chronological vertex updating without sacrificing the computation parallelism.
We evaluate the performance of the proposed hardware accelerator on three real-world datasets. 
The proposed model reduces the computation complexity by $84\%$ and memory accesses by $67\%$ with less than $0.33\%$ accuracy loss. 
Compared with CPU/GPU, our FPGA accelerator achieves $16.4/2.3\times$ speedup in latency and $0.27\%$ improvement in accuracy compared with the state-of-the-art inference algorithm. 
To the best of our knowledge, this is the first work that performs model-architecture co-design on memory-based Temporal Graph Neural Networks.
\end{abstract}


\section{Introduction}

The acceleration of static GNNs operating on {\it structural} and {\it contextual} information to produce node/link {\it embeddings} on static graphs is a burgeoning area of research. Proposed techniques range from algorithmic model optimizations (pruning and compression \cite{channelpruning,Xu2020Dynamically,Wang_2021_CVPR}) to dedicated hardware accelerators \cite{graphact, yan2020hygcn, geng2020awb}. These 
efforts achieve massive task parallelism, allowing static GNNs to be deployed accurately and efficiently on large-scale graphs for a variety of 
applications (recommender systems \cite{ying2018graph,zhang2019star}, knowledge graph reasoning \cite{jin2020recurrent,sedyt}, fraud detection \cite{xianyu} etc.).

However, most real world graphs also contain {\it temporal} information that is sensitive both to duration (longevity) and chronology. For example,  interactions in social networks are often timestamped, components  of knowledge graphs have duration-limited validity and chronological ordering of graph signals is critical for real-world tasks like fraud detection and event prediction. Thus Temporal GNNs (TGNNs) \cite{tgat,tgn,dysat,evolvegcn,dyrep} that generate node embeddings also capturing (evolving) temporal information have become 
\looseness=-1 
popular.

In production environments, TGNNs are usually used to compute dynamic node embeddings on the upcoming stream of graph signals for downstream tasks. %
However, several unique characteristics of TGNNs make them inefficient for deployment on General Purpose Processors (GPPs). 
First, TGNNs are significantly more compute-intensive. 
In order to accurately capture the evolving nature of temporal neighborhoods, most TGNNs \cite{tgat,tgn,apan,wang2021inductive} rely on a temporal attention mechanism (adopted from Transformer \cite{transformer}) to aggregate features from temporal neighbors along with additional sequence models like RNNs and GRUs. An artifact of this mechanism is that it requires computing additional ``keys'' and ``queries'' for each temporal neighbor (more than 2$\times$ the number of operations than a mean or max pooling aggregator). 
Second, graph signals can appear asynchronously, at varying rates. Temporal neighbor sampling and vertex information updates associated with these signals lead to intrinsic sequential dependencies which require the system to process small batches. 
This raises the issue of implementation platform. Current TGNN implementations are mostly GPU focused, where the coarse-grained parallelism usually leads to significantly worse performance on small versus large batches.
Further, both latency {\it and} throughput of processing these signals is important. %
State-of-the-art works like 
\cite{apan} hide latency by offlining part of the message passing process. This requires exponential amount of extra memory to cache  intermediate results making it hard to scale to large dynamic graphs while also not decreasing computational complexity.

We believe that while algorithmic optimizations per se are useful in partially solving the above challenges, a more holistic approach that also leverages the fine-grained parallelism,  low-latency on-chip memory (for customized memory access patterns), and  high density resources (programmable DSPs for customized data paths) of hardware platforms such as FPGAs can provide a superior overall 
\looseness=-1
solution.

In this paper, we present a model-architecture co-design for high-performance TGNN inference on FPGA. To the best of our knowledge, this is the first such co-design framework that jointly optimizes both throughput and latency, consisting of a suite of algorithmic model optimizations to solve computational and memory bottlenecks imposed by model constraints and carefully mapped to maximize FPGA architectural support for performance acceleration.
We begin with an analysis of the computation processes of general memory-based TGNNs along with a case study evaluating computation-communication characteristics. Based on the identified bottlenecks, we perform both algorithmic and hardware-specific optimizations to make TGNN inference computationally tractable with negligible accuracy loss. We design and implement our hardware accelerator on two different FPGAs and demonstrate high throughput and low latency TGNN inference with negligible accuracy loss on real-world datasets.
We summarize our main contributions \looseness=-1 below:
\begin{itemize}
    \item 
    We propose a light-weight temporal attention mechanism and a related neighbor pruning strategy which greatly reduce the computation and memory accesses at inference. We design a knowledge distillation setup to train our simplified models to ensure comparable accuracy.
    \item To better leverage the FPGA hardware, we propose FPGA-specific optimizations that replace the temporal sampler with a FIFO-based hardware sampler and replace the time encoder with a Look-up-Table  (LUT) based time encoder. We use a hardware-based mechanism to rapidly maintain the vertex information up-to-date. We design and implement a hardware accelerator using techniques including batching, pipelining, and prefetching to achieve massive computation parallelism. 
    \item We propose a predictive performance model for our hardware accelerator to estimate the performance based on algorithm parameters, design configurations, and memory characteristics.
    \item We implement the proposed design on state-of-the-art FPGA platforms Xilinx Alveo U200 and Xilinx ZCU104. Compared with the baseline, our hardware accelerators on ZCU104/U200 achieve $2.00/5.04\times$ improvement in latency and $2.46/8.81\times$ improvement in throughput compared with the CPU/GPU baselines.
\end{itemize}
\section{Temporal Graph Neural Networks}
\label{sec:tgnngeneral}

\begin{table*}[t]
    \centering
    \caption{Number (\#) and percentages (\%) of thousands of MEMs (kMEM) and thousands of MACs (kMAC) and the average execution time on CPU and GPU per dynamic node embedding.}
    \setlength{\tabcolsep}{2mm}
    \begin{tabular}{r|cc|cc|ccc||cc|cc|ccc}
         & \multicolumn{7}{c||}{Wikipedia} & \multicolumn{7}{c}{Reddit}\\
         & \multicolumn{2}{c|}{kMEM} & \multicolumn{2}{c|}{kMAC} & \multicolumn{3}{c||}{Exec. Time (ns)} & \multicolumn{2}{c|}{kMEM} & \multicolumn{2}{c|}{kMAC} & \multicolumn{3}{c}{Exec. Time (ns)}\\
         & \# & \% & \# & \% & 1 Thread & 32 Threads & GPU & \# & \% & \# & \% & 1 Thread & 32 Threads & GPU\\
        \toprule 
        sample & 0.0 & 0.3\% & 0 & 0\% & 9 & 9 & 8 & 0.1 & 1.1\% & 0 & 0\% & 11 & 9 & 8\\
        memory & 5.2 & 91.4\% & 48.4 & 6.0\% & 273 & 40 & 8 & 5.2 & 90.7\% & 48.4 & 6.0\% & 198 & 47 & 9\\
        GNN & 0 & 0\% & 703.5 & 93.6\% & 296 & 33 & 4 & 0 & 0\% & 703.5 & 93.6\% & 297 & 31 & 3\\
        update & 0.5 & 8.3\% & 0 & 0\% & 23 & 21 & 19 & 0.5 & 8.2\% & 0 & 0\% & 27 & 25 & 15\\
        \midrule
        total & 5.7 & 100\% & 751.9 & 100\% & 601 & 103 & 39 & 5.8 & 100\% & 751.9 & 100\% & 533 & 112 & 35\\
    \end{tabular}
    \label{tab:basecomp}
\end{table*}

Given a dynamic graph $\mathcal{G}(\mathcal{V},\mathcal{E})$ where  nodes and edges are associated with timestamps representing their appearance/disappearance, the TGNN inference problem aims at encoding contextual, structural, and temporal information of nodes at specific times into dynamic node embeddings and maintaining up-to-date node memory. Note that TGNN inference is distinct from the inference (link prediction etc.) carried out by downstream tasks. In this work, we choose to use the memory-based TGNNs which achieve state-of-the-art accuracy as the baseline models to optimize. We briefly describe the memory-based TGNN inference process below.

For each node $v$, memory-based TGNNs maintain a node memory vector $\bm s_v$  that summarizes its status. A message $\mbox{MSG}(m)$ is generated whenever a  graph signal $m$ related to $v$ occurs. Multiple related messages 
from $v$'s receptive field (of neighbors) $\mathcal{N}(v)$, are combined  by an aggregator function $\mbox{AGGR}$. These are then used to update $\bm s_v$ via an update function $\mbox{UPDT}$. $\bm s_v$ is then fed as an input feature into an attention-based GNN to generate the output node embedding at the current timestamp $\bm h_v$.

\begin{equation}
    \label{eq:memupdt}
    \bm s_v=\mbox{UPDT}(\bm s_v,\mbox{AGGR}(\mbox{MSG}(\{\bm m, \bm m\in\mathcal{N}(v)\})))
\end{equation}
\begin{equation}
    \label{eq:generalgnn}
    \bm h_v=\mbox{GNN}(\mathcal G_v, \{\bm s_u,u\in\mathcal G_v\}),
\end{equation}

\noindent where $\mathcal G_v$ denotes $v$'s supporting temporal neighbors (sampled from the past neighbors of $v$). 

The process in Equation (\ref{eq:memupdt})-(\ref{eq:generalgnn}) is computed both at training and inference.
Since dynamic node labels are difficult to acquire and hardly provide enough information on evolving graphs, TGNNs are usually trained by self-supervision from the temporal edges. 
Under this setup, the temporal edges which need to be predicted by an external downstream edge classifier are itself fed to the TGNN models as input during training, creating an ``information leak'' problem \cite{tgn}.
To solve this, the TGNN {\it caches} the messages of the current graph signals rather than directly use them as input. 
When a new graph signal arrives, the $\mbox{UPDT}$ function takes input from the cached messages instead of using the message of the current signal to update the node memory.
Therefore, at inference, to be consistent with the training phase, we need to first update the node memory by the cached messages (Equation (\ref{eq:memupdt})) before aggregating from them (Equation (\ref{eq:generalgnn})).

\subsection{Inference Performance Metrics}
\label{subsec:performance-metric}
When deployed for real-world applications, TGNN-based systems usually operate on upcoming graph signals in batches, formed either by fixed number of graph signals or by the graph signals in fixed time windows. To quantify the performance of TGNN inference, we formally define our evaluation metrics of throughput and latency. Since most existing datasets only have new edges as graph signals, we define the throughput as the number of new edges that can be processed per second.
Defining the execution time as the total time to generate the dynamic node embeddings and maintain the node memory up-to-date, we define throughput as
\begin{equation}
    \text{TGNN throughput } (\text{E/s})=\frac{\# \text{ of new edges}}{\text{execution time (s)}},
\end{equation}
We define TGNN inference latency as the time to output the required dynamic node embeddings once we receive a batch of graph signals.
We follow the general setup in \cite{tgn,apan,dyrep} for processing incoming graph signals, 
where temporal dependencies are ignored for nodes in the same batch while the node memory and the cached messages are updated in order. 
For the graph signals in one batch, we perform one forward propagation and generate the corresponding dynamic node embeddings for {\it all} the involved nodes, which is the common inference scenario for many applications.
For example, a fraud detection application would like to frequently examine {\it all users} involved in newly appearing transactions. 
Our goal is to increase the throughput and decrease the latency for TGNN inference while maintaining similar accuracy.

\subsection{Case Study: Memory-Based TGNN Inference}
\label{sec:case}

In this subsection, we perform a detailed case study to analyze the computation complexity, memory accesses, and runtime profiling of the memory-based TGNN inference. TGN \cite{tgn} provides a general framework for node memory-based TGNNs and benchmarks the performance with different $\mbox{UPDT}$ and $\mbox{GNN}$ functions. Among these memory-based TGNN variants, 1-layer attention-based GNN with GRU memory updater (TGN-attn) has the highest accuracy to complexity ratio. Hence, we focus on optimizing the performance of TGN-attn in this work. Nevertheless, our proposed optimizations also apply to other TGNNs.

In TGN-attn, when a graph signal of new interaction between nodes $i$ and $j$ appears, two messages are generated 

\begin{align}
    \bm m_i&=\bm s_i||\bm s_j||\bm f_{e} || \Phi(\Delta t)\label{eq:message1} \\
    \bm m_j&=\bm s_j||\bm s_i||\bm f_{e} ||\Phi(\Delta t) \label{eq:message2}
\end{align}
where $||$ denotes concatenation, $\bm f_{e}$ is the edge feature vector, and $\Delta t$ the time difference between the most recent node memory $\bm s$ and the timestamp of the graph signal. $\Phi(\cdot)$ is a time encoder similar to the positional encoder in Transformer with two learnable vectors $\bm\omega$ and $\bm\phi$

\begin{equation}
    \Phi(\Delta t)=\cos(\bm\omega\Delta t+\bm\phi)
    \label{eq:time-encoding-formula}
\end{equation}
All Messages to a node in a batch are aggregated with the "Most-Recent" aggregator which simply keeps the most recent message of each temporal node. The aggregated message $\overline{\bm m_i}$ is then sent to a GRU cell which updates the node memory $\bm s_i$  using the previous node memory as hidden state and aggregated messages as input features, as shown below.

\begin{align}
    \bm r_i&=\sigma(\bm W_{ir}\overline{\bm m_i}+\bm b_{ir}+\bm W_{hr}\bm s_i+\bm b_{hr}) \label{gru:ri}\\
    \bm z_i&=\sigma(\bm W_{iz}\overline{\bm m_i}+\bm b_{iz}+\bm W_{hz}\bm s_i+\bm b_{hz}) \label{gru:zi}\\
    \bm n_i&=\tanh(\bm W_{in}\overline{\bm m_i}+\bm b_{in}+\bm r_i(\bm W_{hn}\bm s_i+\bm b_{hn})) \label{gru:ni}\\
    \bm s_i&=(1-\bm z_i)\bm n_i+\bm z_i\bm s_i \label{gru:si}
\end{align}
where $\bm W$ and $\bm b$ denote learnable weights and biases. The updated node memory is now combined with the static node features $\bm f_i$ and fed into the attention aggregator. 
Since the node memory already contains the status of the nodes in the past, the attention aggregator can focus on recent graph signals. A fixed amount of most recent temporal neighbors $j\in\mathcal{N}_i$ are sampled and fed to the attention aggregator 

\begin{align}
    \bm f'_i&=\bm s_i+\bm W_s\bm f_i+\bm b_s\label{eq:tatt}\\
    \bm q&=\bm{W_q}\left[\bm f'_i||\Phi(0)\right]+\bm{b_q}\label{eq:tattq}\\
    \bm K&=\bm{W_k}\left[\bm f'_j||\bm e_{ij}||\Phi(\Delta t))\right]+\bm{b_k}\label{eq:tattk}\\
    \bm V&=\bm{W_v}\left[\bm f'_j||\bm e_{ij}||\Phi(\Delta t))\right]+\bm{b_v}\label{eq:tattv}\\
    \bm h_i&=\textup{softmax}\left(\frac{\bm q\bm K^\textup{T}}{\sqrt{|\mathcal{N}_v|}}\right)\bm V\label{eq:attagg}
\end{align}
where the attention aggregator performs vector-vector multiplication between the queries $\bm q$ and keys $\bm K$ to determine attention scores $\bm h_i$. 
After computing the required dynamic node embeddings in each batch, we obtain the updated node memory $\{\bm s_i\}$ and the messages $\{\bm m_i\}$ of the involved nodes $\{i\}$. We update the node memory and cache the messages in a global copy which is usually stored in the external memory.

For analysis, we divide the whole process into four parts: sample, memory, GNN, and update. The sample part accesses the dynamic graph and samples most recent temporal neighbors. The memory part aggregates the messages and computes the updated node memory. The GNN part applies attention aggregator to generate the dynamic node embeddings. The update part writes the updated node memory back and updates the cached messages. We calculate the number of MEMory accesses (MEMs) (assuming the learnable parameters are stored on-chip) and Multiplication and ACcumulations (MACs) in each part on two popular dynamic graphs Wikipedia and Reddit \cite{tgat}. 
For each node, we follow the setup in TGN \cite{tgn} and sample 10 most recent neighbors from all temporal neighbors as the supporting nodes. We run an optimized version of the open-sourced code \footnote{\url{https://github.com/twitter-research/tgn}}, which replaces the inefficient python loops with batched operations, on 1) a single thread in the Intel Xeon Gold 5120 CPU 2) 32 threads on the same CPU 3) an Nvidia Titan Xp GPU and measure the execution time of the four parts. 
Table \ref{tab:basecomp} shows the complexity and execution time in the four parts. The memory accesses are primarily in the memory part to access the messages and edge features. The computation is dominated by the GNN part to aggregate from selected temporal neighbors. For a serial processor (1CPU), the bottleneck is the computation in the GNN part. For highly parallel machines (32 CPU threads and GPU), the bottleneck lies in the memory part that accesses the memory and aggregate the messages. Although the number of memory updates is not enormous in the update part, due to the need to maintain the sequential order, it still becomes the bottleneck when executed on a highly-paralleled machine with complex cache hierarchy.
\section{Model-Architecture Co-Design}
\label{sec:Model-Architecture Co-Design}
We use our case study on memory-based TGNN inference in Section \ref{sec:case} to identify {\it three key points} in designing an optimized inference system:

\begin{enumerate}
    \item GNN computation accounts for more than 80\% of the total time and is the bottleneck on a single CPU core with more than 50\% of the time spent on computing the attention scores (Equation (\ref{eq:tattq}) and (\ref{eq:tattk})). It is linear in the number of supporting temporal neighbors.
    \item The time encoding maps a scalar time interval to a vector which is further multiplied with weight matrices $\bm W_{ir}$, $\bm W_{iz}$, $\bm W_{in}$, $\bm W_q$, $\bm W_k$, and $\bm W_v$. These vector-matrix multiplications can be removed if we can reverse the computation order.
    \item On a massively parallel architecture, the key bottleneck is fetching and updating the node memory and messages of the supporting nodes from and to external memory.
\end{enumerate}
Based on these key points, we propose a model-architecture co-design by exploiting FPGA-specific features. FPGAs consist of massive on-chip memory -- Block RAMs (BRAMs) and Ultra RAMs (URAMs) that allow fast random memory access in high bandwidth. The built-in DSPs can perform large number of arithmetic operations in each cycle. Based on these features, we propose a simplified temporal attention mechanism (Section \ref{sec:simatt}) and a temporal neighbor pruning strategy (Section \ref{sec:neighprune}) which reduce the execution time in DSPs and enable data pre-fetching.
In addition, the DSPs of FPGAs can be programmed into computation arrays that enable fine-grained parallelism for batched processing.  
We replace the time encoding and the subsequent vector-matrix multiplications with look-up tables (LUTs) (Section \ref{sec:timelut}) which are stored in the programmable on-chip memory of FPGAs for fast accesses. Besides, the on-chip memory of FPGAs can be programmed into customized cache structures (Section \ref{sec:hw-modules}) which facilitate fast updates to vertex data.

\subsection{Simplified Temporal Attention}
\label{sec:simatt}

As seen in Equation~(\ref{eq:attagg}), the traditional temporal attention mechanism requires vector-vector multiplication among neighbors which consumes a lot of DSPs on FPGAs. We note that temporal neighbors in dynamic graphs can be naturally ordered by timestamp.
We leverage this unique characteristic to design a simplified attention aggregator that operates on fixed-length lists of $n$ timestamp-sorted (not necessarily unique) temporal neighbors.
Given a node $u$ at timestamp $t^u$ with $n$ sorted temporal neighbors 
at respective timestamps $t^{v_0}\leq t^{v_1}\leq\cdots\leq t^{v_{n-1}}$, 
we compute its attention score 
\looseness=-1 as

\begin{equation}
    \bm\alpha'(u)=\textup{Softmax}(\bm a+\bm W_t\bm{\Delta t}^u)
    \label{eq:catt}
\end{equation}
where $\bm a$ is a learnable constant attention vector shared among all  nodes and $\bm W_t$ is a learnable weight matrix that maps node-specific time difference $\bm{\Delta t}^u=[t^u-t^{v_0},\cdots,t^u-t^{v_{n-1}}]$ to respective offsets of the attention logits. 
The intuition behind Equation (\ref{eq:catt}) is that on a dynamic graph, attention scores should be sensitive to chronology of neighbors. 
Since each node $u$ has a specific neighbor interaction frequency, we use this node-specific offset to produce its attention score. 
Our simplified attention mechanism eliminates the vector-vector multiplication operations (Equation (\ref{eq:attagg})) which dramatically saves the DSP usage on FPGAs.

To learn $\bm a$ and $\bm W_t$, we apply a simplified knowledge distillation \cite{hinton2015distilling} setup under which we train  {\it student models} (with our simplified temporal attention aggregators) under both  self-supervision from temporal edges and supervision from a teacher model with the vanilla temporal attention aggregator. We add an additional soft cross-entropy loss $l_a$ between the simplified attention logits $\overline{\bm\alpha}'=\bm a+\bm W_t\bm{\Delta t}$ and the vanilla attention logits $\overline{\bm\alpha}$ to encourage the student models to mimic the behaviour of the teacher model 

\begin{equation}
    l_a = -\sum_{v}\textup{Softmax}(\overline{\bm\alpha}'(v)/T)\cdot\textup{Softmax}(\overline{\bm\alpha}(v)/T)
\end{equation}
where $T$ is the temperature that controls how much the student model learns from the teacher model.

\subsection{Temporal Neighbor Pruning}
\label{sec:neighprune}

The Transformer attention mechanism shown in Equation (\ref{eq:tatt})-(\ref{eq:tattv}) computes the attention scores after the computation of the keys $\bm K$ and queries $\bm Q$. In contrast, our simplified temporal attention mechanism computes the attention scores only using the time difference $\bm{\Delta t}$ of the temporal neighbors as the input. This allows models to quickly determine which temporal neighbor is more important before fetching the hidden features from them all. Although the amount of computation and number of memory accesses are the same for each temporal neighbor, the neighbors with higher attention scores contribute more to the output hidden features, which naturally leads to our attention score-based temporal neighbor pruning strategy. Under the simplified attention mechanism where only the values $\bm V$ needs to be computed, performing temporal neighbor pruning directly leads to a linear reduction in computation and memory accesses. For a given pruning budget (number of temporal neighbors to aggregate from), after computing the logits of the simplified attention scores, we apply softmax function only on the temporal neighbors with top logit values and compute their $\bm V$. 

\subsection{Time Encoding Look-Up-Table}
\label{sec:timelut}

\begin{figure}[t]
    \centering
    \vspace{-0.3cm}
    \input{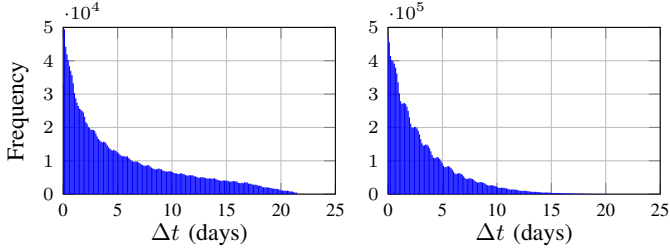}
    \vspace{-0.3cm}
    \caption{Frequency of input $\Delta t$ of the time encoder on the Wikipedia (left) and Reddit (right) datasets.}
    \label{fig:timefreq}
\end{figure}

The time encoder in Equation (\ref{eq:time-encoding-formula}) maps scalar time frames to vectors. These vectors are later multiplied with the weight matrices in the GRU memory updater and the GNN aggregator. This whole process accounts for around 30\% of the total computation in the TGNN model with our simplified attention mechanism. This can be completely avoided if the computation order is reversed and the vector-matrix multiplication is pre-computed. However, the time encoding process involves a nonlinear trigonometric function which does not permit pre-computation. To solve this problem, we replace the time encoding process with LUT operations which transforms the nonlinear operations to piece-wise linear ones. We analyze the input $\Delta t$ of the time encoder and observe that it follows the power law where most inputs are close to 0. Based on the intuition that the output vectors should distinguish different length of time frames, we divide the range of the input $\Delta t$ to 128 intervals with equal number of $\Delta t$ occurrences in each interval. The output time encoding vectors in each interval are stored in one entry in the LUT, which is learned in the training process. At inference, we pre-compute the product of each entry in the LUTs with the weight matrices and store them in the fast on-chip memory. Our LUT-based time encoder can directly output the hidden features after weight application for any given time frame within 1 clock cycle.

\begin{figure*}[h]
    \centering
    \vspace{-0.3cm}
    \input{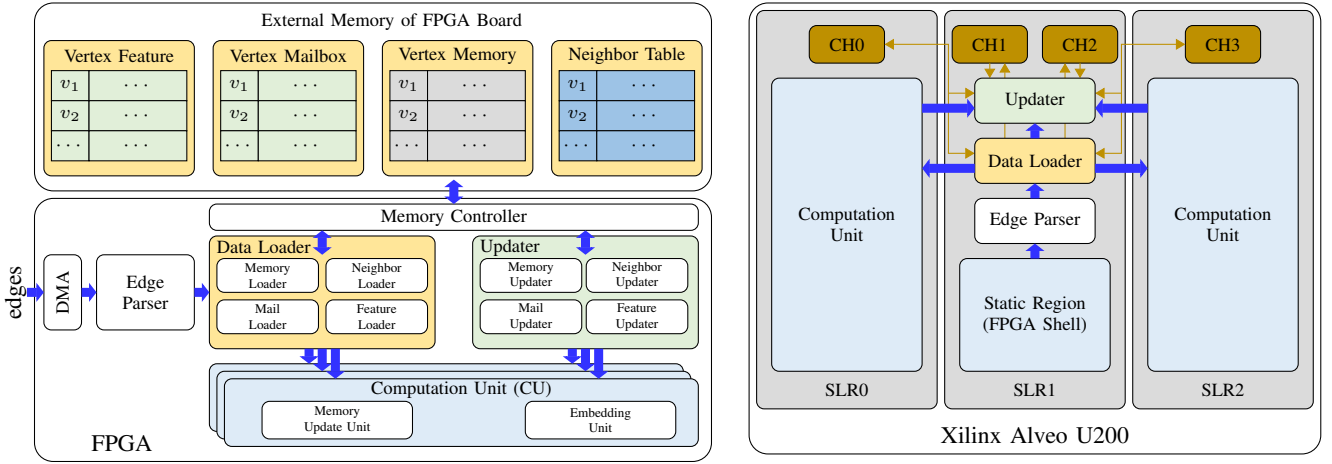}
    \vspace{-0.3cm}
    \caption{The principled hardware architecture for TGNN on FPGA platform (Left). Mapping of the architecture on Xilinx Alveo U200 platform (Right).}
    \label{fig:high-level-arch}
\end{figure*}

\section{Hardware Mapping and Optimizations}

 \textcolor{black}{ Figure \ref{fig:high-level-arch} illustrates the overview of the proposed architecture that executes the inference process of TGNN (Algorithm \ref{algo:inference-process}).  The proposed architecture consists of the following parts:
\begin{itemize}
    \item \textbf{Input}: The data packets of the new edges are sent to FPGA accelerators through the Direct Memory Access (DMA) unit. 
    \item \textbf{Graph Storage} The FPGA board consists of an external DDR memory and an FPGA chip. The external DDR memory stores various vertex information, including the Vertex Mailbox $\{\bm{m}_{v}: v \in \mathcal{V} \}$, the Vertex Memory Table $\{\bm{s}_{v}: v \in \mathcal{V} \}$, the vertex feature vectors  $\{\bm{f}_{v}: v \in \mathcal{V} \}$, and the Vertex Neighbor Table $\{\mathcal{N}_{mr}(v): v \in \mathcal{V}\}$.
    \item \textbf{Accelerator}: The proposed hardware accelerator consists of four parts -- the Edge Parser, the Data Loader, the Updater, and the Computation Units. The memory controller handles the data transmissions between the external memory and the accelerator. The Edge Parser receives the new edges from the host processor through DMA and parses the raw information of the new edges. The Data loader loads the required inputs from external memory. The Updater ensures the chronological order of the updated vertex information and sends the updated vertex information back to external memory. The Computation Units (CUs) perform the key computation stages of TGNN inference where each CU has its own Memory Update Unit (MUU) to generate the updated vertex memory and an Embedding Unit (EU) to generate the updated vertex embedding.
\end{itemize} }

\noindent \textcolor{black}{ \textbf{Runtime}: At runtime, the accelerator first receives a new batch of edges (line 2 of Algorithm \ref{algo:inference-process}). Then, the MUU calculates the vertex memory of involving vertices and the Updater updates the vertex memory and vertex cache messages (line 3-8 of Algorithm \ref{algo:inference-process}) in the external memory of FPGA board. After that, the EU calculates the vertex embeddings for the downstream tasks (line 9-11 of Algorithm \ref{algo:inference-process}) and the vertex neighbor table is updated based on the new edges (line 12-14 of Algorithm \ref{algo:inference-process}). To improve the overall throughput, we adapt batching, fine-grained task pipelining and prefetching to fully exploit the parallelism (Section \ref{subsec:dataflow-optimizations}).}

\subsection{Data Structure}
We represent each edge in a dynamic graph as $e(src, dst, \bm f_{e}, t_{e})$ where $src$, $dst$, $\bm{f}_{e}, t_{e}$ denote the source index, destination index, feature vector, and timestamp, respectively. Vertex information consists of the Vertex Mailbox (cached messages), the Vertex Memory Table, and the Vertex Neighbor Table. Each row of the Vertex Mailbox contains the most recent message $\bm m_{v}$ of a vertex. Each row of the Vertex Memory Table contains the most recent vertex memory $\bm s_{v}$ of a vertex. Each row of the Vertex Neighbor Table contains the indices of the most recent $mr$ neighbors $\mathcal{N}_{mr}(v)$ of a vertex.  

\begin{algorithm}[t]
\small
 \caption{Inference process of Memory-based TGNN on the proposed accelerator}
 \begin{algorithmic}[1]
 \renewcommand{\algorithmicrequire}{\textbf{Input:}}
\renewcommand{\algorithmicensure}{\textbf{Output:}}
 \REQUIRE  Graph $\mathcal{G}(\mathcal{E},\mathcal{V})$; A edges stream $\mathcal{E}_{\text{new}}$ that incoming edges follow the chronological order; Old vertex memory $\{\bm s_{v}: v \in \mathcal{V} \}$; Old cached messages $\{\bm m_{v}: v \in \mathcal{V} \}$; Vertex feature vectors  $\{\bm{f}_{v}: v \in \mathcal{V} \}$; Edge feature vectors $\{\bm f_e:e\in\mathcal{E}\}$;  
 \STATE \COMMENT{ \textcolor{blue}{\% Process batches of new edges in chronological order \%}}
     \FOR{each batch $\{e(u,v,\bm f_{e},t_{e})\}\in\mathcal{E}_{\text{new}}$}
        \STATE \COMMENT{\textcolor{blue}{\% update vertex memory \%}}
        \STATE $\{{\bm s}_{u}\}=\{\text{UPDT}(\bm m_{u},\bm s_{u},t_{e})\}$
        \STATE $\{\bm {s}_{v}\}=\{\text{UPDT}(\bm m_{v},\bm s_{v},t_{e})\}$
        \STATE \COMMENT{\textcolor{blue}{\% update cached messages \%}}
        \STATE $\{\bm {m}_{u}\}=\{\bm s_{u}||\bm s_{v}||\bm f_{e}||\Phi(t_{e})\}$
        \STATE $\{\bm {m}_{v}\}=\{\bm s_{v}||\bm s_{u}||\bm f_{e}||\Phi(t_{e})\}$
        \STATE \COMMENT{\textcolor{blue}{\% compute output embeddings \%}}
        \STATE $\{\bm{h}_{u}\}=\{\text{GNN}((\bm s_u,\bm f_{u},t_e), \{(\bm s_{z},\bm f_{z},t_z), z\in\mathcal{N}(u)\})\}$
        \STATE $\{\bm{h}_{v}\}=\{\text{GNN}((\bm s_v,\bm f_{v},t_v), \{(\bm s_{z}, \bm f_{z},t_z),z\in \mathcal{N}(v)\})\}$
        \STATE \COMMENT{\textcolor{blue}{\% update vertex neighbors \%}}
        \STATE $\{\mathcal{N}(v)=\text{UpdateNeighbor}(\mathcal{N}(u),v)\}$
        \STATE $\{\mathcal{N}(u)=\text{UpdateNeighbor}(\mathcal{N}(v),u)\}$
    \ENDFOR 
  \end{algorithmic} 
 \label{algo:inference-process}
 \end{algorithm}

\subsection{Hardware Modules}
\label{sec:hw-modules}

In the proposed accelerator, the $\mbox{UPDT}(\cdot)$ function is implemented as a GRU (Equation (\ref{gru:ri}), (\ref{gru:zi}), (\ref{gru:ni}), (\ref{gru:si})) and the $\mbox{GNN}(\cdot)$ function is implemented as a 1-layer GNN with attention mechanism (Equation (\ref{eq:catt})). The GRU is mapped to the MUU and the 1-layer GNN is mapped to the EU. The CU performs batched execution for vertex information updating. When MUU is ready to receive new inputs, the vertex messages $\{\bm m_{v}:v \in \mathcal{V}_{b}\}$ and vertex memory $\{\bm s_{v}:v \in \mathcal{V}_{b}\}$ of $N_{b}$ vertices $\mathcal{V}_{b}$ ($|\mathcal{V}_{b}| = N_{b}$) are sent to MUU for memory updating. The updated vertex memory are sent to EU to generate vertex embedding.

\noindent \textbf{Memory Update Unit}: In GRU, there are an Update Gate (Equation (\ref{gru:ri})), a Reset Gate (Equation (\ref{gru:zi})), a Memory Gate (Equation (\ref{gru:ni})) ,and a Merging Gate (Equation (\ref{gru:si})). The Update Gate, Reset Gate, and Memory Gate involve matrix multiplication between the vertex messages (Equation (\ref{eq:message1}), (\ref{eq:message2})) of $N_{b}$ vertices and the weight matrices. For each one of the three gates, a Multiply-Accumulate Array of size $S_{g}\times S_{g}$ is implemented for efficient matrix multiplication. 
The four gates are connected through the on-chip FIFO and their execution is pipelined to achieve task-level parallelism.

\noindent \textbf{Embedding Unit}: The EU performs 1-layer message passing with an Attention Module (AM) to calculate the attention scores of the neighbors $\{\alpha(u):u\in \mathcal{N}_{mr}(v) \}$, a Feature Aggregation Module (FAM) to perform feature aggregation: $\bm{h}_{v} = aggregate\{\alpha(u)\cdot \bm{s}_{u}:u\in \mathcal{N}_{mr}(v) \cup \{v\} \}$ and a Feature Transformation Module (FTM) to perform feature transformation $\bm{h}_{v} = transform(\bm{h}_{v}, \bm{s}_{v}, \bm{W}_{v})$. The FAM uses the multiply-add tree-based design to aggregate information from the neighbors. FTU performs multiplication of the aggregated vertex memory vectors and the weight matrix which is also implemented as a Multiply-Accumulate Array.

\begin{figure}[b]
    \centering
    \vspace{-0.3cm}
    \input{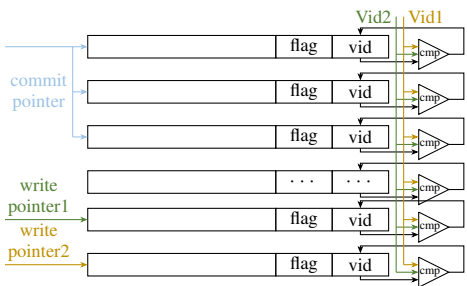}
    \vspace{-0.3cm}
    \caption{Updater using a fully-associative cache with rotating pointers ($N_{cu} = 2$).}
     \label{fig:detail-architecture-updater}
\end{figure}

\begin{figure*}[t]
    \centering
    \vspace{-0.3cm}
    \input{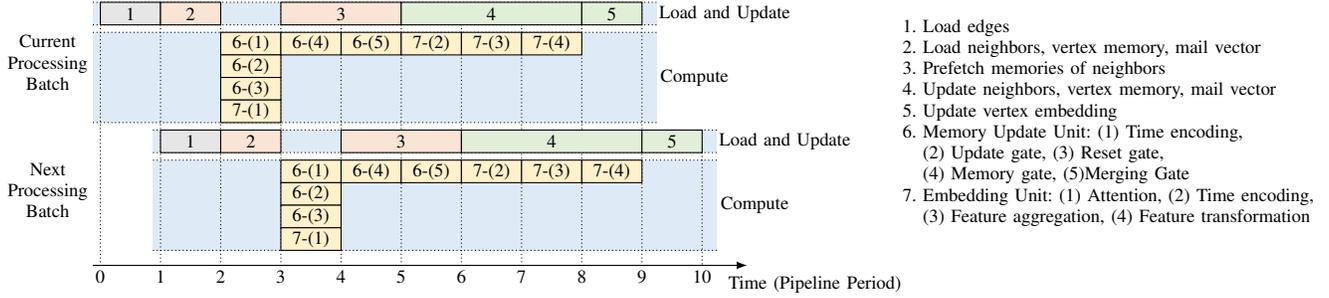}
    \vspace{-0.1cm}
    \caption{Task scheduling.}
     \label{fig:Task-scheduling}
\end{figure*}

\noindent \textbf{Updater}: The function of Updater is to \circled{1} receive the vertex information from the computation units, \circled{2} write the vertex information back to the external memory, \circled{3} ensure the chronological order of the updated vertex information, and \circled{4} eliminate the redundant vertex updating. To ensure the chronological order, the new edges are assigned to the CUs in Round-Robin style. Similarly, the Updater receives the updated vertex information from the CUs in the same Round-Robin order. Figure \ref{fig:detail-architecture-updater} shows the diagram of the Updater. The Updater is organized as a fully-associative cache with rotating pointers. Each cache line stores one vertex information (message/memory/neighbors), the index of the vertex, and a flag bit that indicates whether the current cache line is valid. When the Updater receives the vertex information from multiple CUs concurrently, the chronological order of the multiple vertex information is ensured by the relative position of the write pointers. Each write pointer points to the write position of a CU. The commit pointer scans multiple consecutive cache lines at a time to check if a valid cache link can be sent back to the external memory. When multiple new updated vertex information are received by the Updater, their vertex indices are compared with vertex index ($vid$) of each cache line. If an uncommitted cache line has the same vertex index with new vertex information, this uncommitted cache line will be invalidated.

\noindent \textbf{Multi-die Implementation}: Advanced FPGA platforms are usually integrated with multiple dies with limited number of inter-die connections. The right side of the Figure \ref{fig:high-level-arch} demonstrates our multi-die implementation on Xilinx Alveo U200 board. Multiple memory channels are connected to the data loader and updater. The hardware modules are distributed into different dies (Super Logic Regions) with on-chip FIFOs as the interconnection.

\subsection{Dataflow Optimizations}
\label{subsec:dataflow-optimizations}

We exploit three design principles to optimize the overall performance: batching, task pipelining, and prefetching. Figure \ref{fig:Task-scheduling} depicts the detailed task scheduling.

\noindent \textbf{Batching}: In each time interval $T_{p}$, the computation unit groups a set of edges to start the execution. This exploits data parallelism within the CUs.

\noindent \textbf{Task pipelining}: The execution of consecutive processing batches are fully pipelined to improve the overall throughput. To facilitate the pipelined execution, (1) the computation operations and the memory accesses are overlapped, (2) the MMU and EU are pipelined, and (3) the individual hardware modules within MUU and EU are pipelined.  

\noindent \textbf{Prefetching}: $\mbox{GNN}(\cdot)$ requires the vertex memory of the neighbors where loading from external memory leads to large latency. Therefore, in the task scheduling (Feature \ref{fig:Task-scheduling}), the calculation of the attention score (7-(2)) in the EU is executed after loading the timestamps of the neighbors. The EU uses the attention score to obtain the indices of the required neighbors and then prefetches the memory of the neighbors before the MUU finishes execution. Note that the prefetching of vertex memory is promoted by the proposed attention mechanism (Equation \ref{eq:catt}) which does not require the updated vertex memory to obtain the attention scores.

\section{Performance Model}

\begin{table*}[t]
    \centering
    \caption{Accuracy, computation complexity, latency, and throughput (with a single CPU thread) of the original and optimized models. The SAT row denotes our simplified attention mechanism. The LUT row denotes LUT-based time encoder. The NP(L/M/S) rows denotes the neighbor pruning with 6/4/2 neighbors. We show the model with accumulated optimizations row by row.}
    \setlength{\tabcolsep}{1.5mm}
    \begin{tabular}{rr|ccc|cc|cccc|cc|cc}
         & & \multicolumn{3}{c|}{Input Dimensions} & \multicolumn{2}{c|}{kMEM} & \multicolumn{4}{c|}{kMAC} & & & \multicolumn{2}{c}{1 CPU Thread} \\
         & Model & $|\bm v_i|$ & $|\bm e_{ij}|$ & $|\mathcal{N}(v)|$ & \# & \% & \#(GRU) & \#(GNN) & \#(Tot.) & \%(Tot.) & \multicolumn{2}{c|}{AP (difference)} & Thpt. (kE/s) & Speedup\\
        \toprule
        \multirow{6}{*}{\STAB{\rotatebox[origin=c]{90}{Wikipedia}}} & Baseline & 0 & 172 & 10 & 5.7 & 100\% & 48.4 & 703.5 & 751.9 & 100\% & 0.9900 & (-0.0000) & 0.85 & 1$\times$\\
         & +SAT & 0 & 172 & 10 & 5.7 & 100\% & 48.4 & 351.1 & 399.5 & 53.1\% & 0.9821 & (-0.0079) & 1.10 & 1.29$\times$\\
         & +LUT & 0 & 172 & 10 & 5.7 & 100\% & 38.3 & 240.0 & 278.3 & 37.0\% & 0.9891 & (-0.0009) & 1.12 & 1.32$\times$\\
         & +NP(L) & 0 & 172 & 6 & 3.8 & 66.7\% & 38.3 & 156.4 & 194.5 & 25.9\% & 0.9891 & (-0.0009) & 1.71 & 2.01$\times$\\
         & +NP(M) & 0 & 172 & 4 & 2.9 & 50.9\% & 38.3 & 114.6 & 152.9 & 20.3\% & 0.9887 & (-0.0013) & 2.71 & 3.19$\times$\\
         & +NP(S) & 0 & 172 & 2 & 1.9 & 33.3\% & 38.3 & 72.8 & 111.1 & 14.8\% & 0.9878 & (-0.0022) & 3.22 & 3.79$\times$\\ 
        \midrule
        \multirow{6}{*}{\STAB{\rotatebox[origin=c]{90}{Reddit}}} & Baseline & 0 & 172 & 10 & 5.8 & 100\% & 48.4 & 703.5 & 751.9 & 100\% & 0.9978 & (-0.0000) & 0.92 & 1$\times$\\
         & +SAT & 0 & 172 & 10 & 5.8 & 100\% & 48.4 & 351.1 & 399.5 & 53.1\% & 0.9967 & (-0.0011) & 1.22 & 1.33$\times$\\
         & +LUT & 0 & 172 & 10 & 5.8 & 100\% & 38.3 & 240.0 & 278.3 & 37.0\% & 0.9978 & (-0.0000) & 1.21 & 1.32$\times$\\
         & +NP(L) & 0 & 172 & 6 & 3.9 & 67.2\% & 38.3 & 156.4 & 194.5 & 25.9\% & 0.9971 & (-0.0007) & 1.51 & 1.64$\times$\\
         & +NP(M) & 0 & 172 & 4 & 3.0 & 51.7\% & 38.3 & 114.6 & 152.9 & 20.3\% & 0.9971 & (-0.0007) & 1.93 & 2.10$\times$\\
         & +NP(S) & 0 & 172 & 2 & 2.0 & 34.4\% & 38.3 & 72.8 & 111.1 & 14.8\% & 0.9948 & (-0.0030) & 2.21 & 2.40$\times$\\
        \midrule
        \multirow{6}{*}{\STAB{\rotatebox[origin=c]{90}{GDELT}}} & Baseline & 200 & 0 & 10 & 5.1 & 100\% & 51.2 & 733.8 & 785.0 & 100\% & 0.9623 & (-0.0000) & 1.29 & 1$\times$\\
         & +SAT & 200 & 0 & 10 & 5.1 & 100\% & 51.2 & 371.3 & 422.5 & 53.8\% & 0.9612 & (-0.0011) & 1.83 & 1.42$\times$\\
         & +LUT & 200 & 0 & 10 & 5.1 & 100\% & 41.1 & 260.2 & 301.3 & 38.4\% & 0.9605 & (-0.0018) & 1.85 & 1.43$\times$\\
         & +NP(L) & 200 & 0 & 6 & 3.4 & 66.7\% & 41.1 & 176.6 & 217.7 & 27.7\% & 0.9598 & (-0.0025) & 3.01 & 2.33$\times$\\
         & +NP(M) & 200 & 0 & 4 & 2.5 & 49.1\% & 41.1 & 134.8 & 175.9 & 22.4\% & 0.9596 & (-0.0027) & 3.62 & 2.81$\times$\\
         & +NP(S) & 200 & 0 & 2 & 1.7 & 31.5\% & 41.4 & 93.0 & 134.1 & 17.1\% & 0.9590 & (-0.0033) & 4.43 & 3.43$\times$\\
    \end{tabular}
    \label{tab:compacc}
\end{table*}

In this section, we introduce our accurate performance model which predicts the performance of the proposed model-architecture co-design on a given FPGA. We define the following notations:
\begin{itemize}
    \item Length of feature vector, message vector, memory, neighbor list, and embedding of a vertex are $f_{\text{feat}}$, $f_{\text{mail}}$, $f_{\text{mem}}$, $mr$, and $f_{\text{emb}}$. The size of each data is $Z_{d}$ bytes.
    \item Number of CUs: $N_{cu}$. Computation Parallelism of each Gate in MUU: $S_{g}\times S_{g}$. Computation parallelism of FAM: $S_{\text{FAM}}$. Computation parallelism of FTM: $S_{\text{FTM}}$.
    \item Number of edges processed concurrently in a pipeline stage (size of a processing batch): $N_{b}$ (see Figure \ref{fig:Task-scheduling}).
    \item Frequency of FPGA design: $F_{\text{freq}}$
\end{itemize}

Due to our task scheduling (Figure \ref{fig:Task-scheduling}), the execution of a processing batch of $N_{b}$ edges are divided into 9 stages. The time period $T_{p}$ of a pipeline stage is decided by the longest stage $T_{\text{comp}}^{\text{max}}$ or the time of loading and storing data from/to external memory $T_{\text{LS}}$:
\begin{equation}
    T_{p} = \max(T_{\text{comp}}^{\text{max}}, T_{\text{LS}})
\end{equation}
where
\begin{equation}
\begin{split}
    T_{\text{comp}}^{\text{max}} = \max(&\{T_{6-(i)}:i=1,..,5\}, \\ &\{T_{7-(i)}:i=1,..,4\})
\end{split}
\end{equation}
For example, $T_{6-(1)}$ is the execution latency of the time encoding stage in MUU as shown in Figure \ref{fig:Task-scheduling}.  $T_{\text{comp}}^{\text{max}}$ can be approximated by the dominant terms
\begin{equation}
\begin{split}
    & T_{\text{comp}}^{\text{max}} \approx \frac{1}{F_{\text{freq}}} \cdot \max(\frac{3\cdot N_{b}\cdot f_{\text{mail}} \cdot f_{\text{mem}}}{S_{g}\cdot S_{g}}, \\ & \frac{3\cdot N_{b}\cdot mr \cdot (f_{\text{mem}} + f_{\text{feat}} )}{S_{\text{FAM}}}, \frac{3\cdot N_{b} \cdot  (f_{\text{mem}} + f_{\text{feat}} )\cdot f_{\text{emb}}}{S_\text{FTM}}).
\end{split}
\end{equation}
To drive an expression for $T_{LS}$, we assume the external memory bandwidth is $\alpha(l) \cdot BW$, where $\alpha(l),(0<\alpha(l)\leqslant1)$ is a factor specifying the effective bandwidth when the length of burst data transaction is $l$ \cite{lu2021demystifying} and $BW$ is the peak memory bandwidth between FPGA and external memory. Similarly, $T_{\text{LS}}$ can be approximated as:
\begin{equation}
\begin{split}
    T_{\text{LS}} \approx & \frac{6\cdot N_{b}\cdot f_{\text{mail}} \cdot Z_{d}}{\alpha(f_{\text{mail}}) \cdot BW }  + \frac{ 3\cdot N_{b}\cdot (2+mr)\cdot f_{\text{mem}} \cdot Z_{d}}{\alpha(f_{\text{mem}}) \cdot BW } \\
    & + \frac{3\cdot N_{b}\cdot mr \cdot f_{\text{feat}} \cdot Z_{d} }{\alpha(f_{\text{feat}}) \cdot BW } + \frac{3\cdot N_{b}\cdot  f_{\text{emb}} \cdot Z_{d}}{\alpha(f_{\text{emb}}) \cdot BW }
\end{split}
\end{equation}
When the batch size $N$ is far larger than the processing batch size $N_b$, the maximum throughput and latency can be expressed as:
\begin{equation}
    \begin{split}
        \text{Maximum Throughput} &  \approx  \frac{N_{b}}{T_{p}}\\
        \text{Latency}  & \approx (\beta - 1 + \lceil \frac{N}{N_{b}} \rceil  ) \cdot T_{p}
    \end{split}
\end{equation}
where $\beta$ is the number of pipeline stages in task scheduling.
\begin{table}[b]
\caption{Specifications of various hardware platforms}
\begin{adjustbox}{max width=0.48\textwidth}
    \setlength{\tabcolsep}{1mm}
    \begin{tabular}{c|c|c|c}

        Hardware & \begin{tabular}[c]{@{}c@{}} \# of dies\\ /sockets/boards \end{tabular} & \begin{tabular}[c]{@{}c@{}}Computation \\ Resources Per Die\end{tabular}                                  & \multicolumn{1}{c}{\begin{tabular}[c]{@{}c@{}}External Memory\\  Bandwidth\end{tabular}} \\
        \midrule
        \begin{tabular}[c]{@{}c@{}}  
         Xilinx \\ Alveo U200 \end{tabular} & 3          & \multicolumn{1}{c|}{\begin{tabular}[c]{@{}l@{}}394K LUTs, 2280 DSPs\\ 720 BRAMs, 320 URAMs\end{tabular}} & 77 GB/s  DDR4                                                                                 \\ \midrule
        \begin{tabular}[c]{@{}c@{}}  
         Xilinx \\ ZCU104 \end{tabular}     & 1          & \begin{tabular}[c]{@{}c@{}}230K LUTs, 1728 DSPs\\ 312 BRAMs,  96 URAMs\end{tabular}                      & 19.2 GB/s DDR4   \\  \midrule    \begin{tabular}[c]{@{}c@{}}  
         Dual Intel Xeon \\ Gold 5120 CPU \\ (CPU baseline) \end{tabular}   & 2 &  \begin{tabular}[c]{@{}c@{}}  14 Cores, 28 Threads\\ 2.20 GHz \end{tabular} &  89 GB/s DDR4          \\ \midrule
          \begin{tabular}[c]{@{}c@{}} Nvidia Titan X\\ (GPU baseline)   \end{tabular} & 1 &  \begin{tabular}[c]{@{}c@{}} 3840 CUDA cores \\  1532 MHz  \end{tabular} &   547 GB/s HBM  \\ 
                                 
    \end{tabular}
\end{adjustbox}
\label{tab:specification-fpga}
\end{table}

\section{Implementation and Experimental Results}

\begin{figure*}[t]
    \centering
    \input{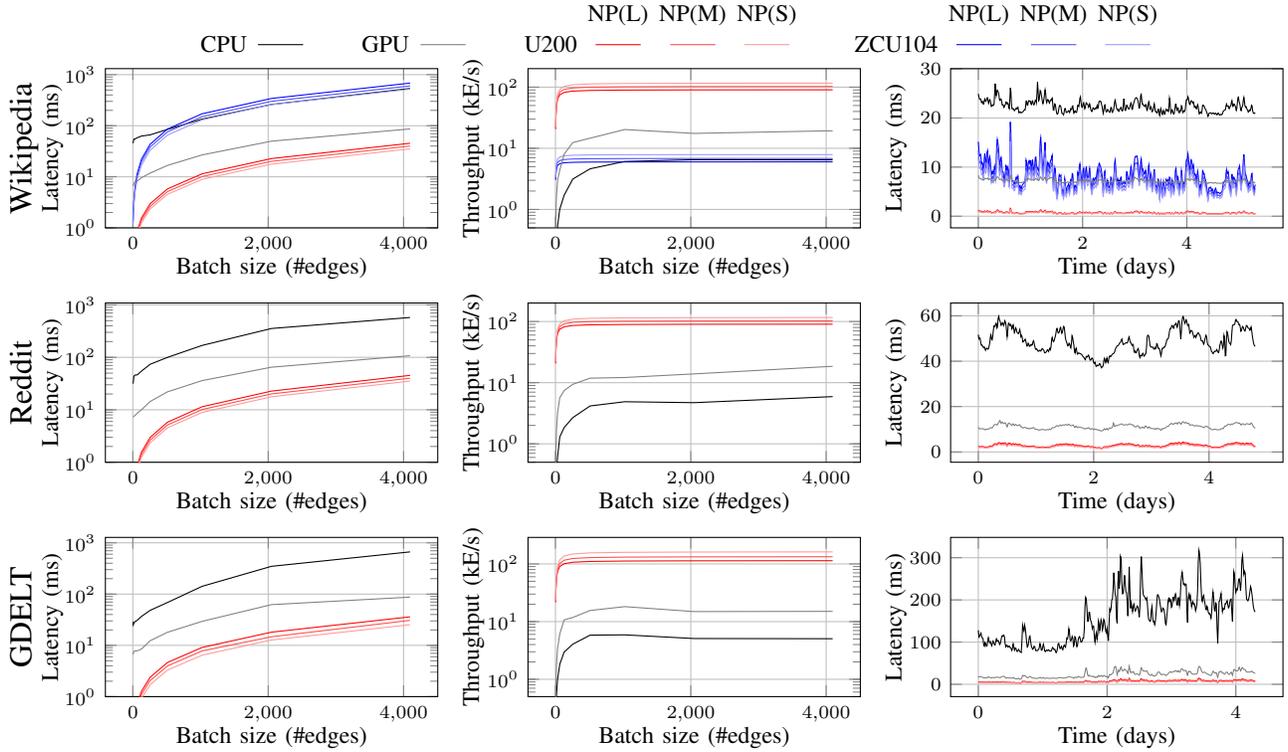}
    \caption{Performance of our hardware accelerator on two FPGAs compared with the baseline on CPU and GPU. The left two plots show the latency and throughput (in log scale) under various batch sizes. The right plots show the real-time inference latency on the test set.}
    \label{fig:main}
\end{figure*}

We evaluate the performance of our proposed model-architecture co-design with the widely-studied temporal link prediction task on three datasets -- Wikipedia \cite{tgat}, Reddit \cite{tgat}, and GDELT \cite{jin2020recurrent}. For the GDELT dataset, we use the pre-trained 200-dimensional node embedding from SeDyT \cite{sedyt} as the input node features. These three datasets cover the dynamic graphs with edge features (Wikipedia and Reddit) and with node features (GDELT). We implement our hardware accelerators on two FPGAs.  Table \ref{tab:specification-fpga} shows the specifications of the hardware platforms used in this work. 
We train three different sizes of the simplified models NP(L/M/S) with 6/4/2 temporal neighbors as the neighbor pruning budgets.
We set the temperature $T$ in the knowledge distillation loss to 1 during training. For the rest of the hyper-parameters, we follow the training setup in the baseline code \cite{tgn}. 

\subsection{FPGA Implementation Results}

\begin{table}[b]
\centering
\caption{Design configurations and Resource utilization}
\setlength{\tabcolsep}{0.9mm}
    \begin{tabular}{c|cccc|cccc|c}
           & \multicolumn{4}{c|}{Design configurations} & \multicolumn{4}{c|}{Resource utilization } & Freq.\\
           & $N_{cu}$ & $S_{g}^2$ & $S_{\text{FAM}}$ & $S_{\text{FTM}}$ & LUT & DSP & BRAM & URAM & (MHz) \\
           \midrule
    U200   &  $2$        &      $8^2$               & $16$         &  $8\times8 $           &  $563$k   &  $2512$     &    $1415$     & $448$    &          250 \\  
    ZCU104 &      $1$     &            $4^2$  &   $8$      &      $4\times 4$      &     $125$k       &  $744$     &    $240$   &    0   &          125 \\ 
    \end{tabular}
\label{tab:config-and-utilization}
\end{table}

We implement our design on two state-of-the-art FPGA platforms -- Xilinx Alveo U200 and Xilinx ZCU104\footnote{\url{https://github.com/zjjzby/TGNN-FPGA-IPDPS2022}}. The accelerators are developed using Xilinx High-level Synthesis (HLS). HLS is a pragma-directive programming language that allows user to develop the accelerator design using C/C++. Alveo U200 is a cloud-based FPGA board while ZCU104 is an embedded FPGA board with a built-in ARM processor. We use ZCU104 to demonstrate that the proposed design can be deployed on light-weight embedded platforms, which is useful for applications on edge devices such as Internet of Things (IoTs). 
We set the commit pointer in the Updater to scan 3 consecutive cache lines each cycle.
A BRAM has the size of 36K bits and an URAM has the size of 288K bits on both FPGAs. With the IEEE float32 data format, each multiplier consumes 3 DSPs while each accumulator consumes 2 DSPs. The design configurations and resource utilization of the accelerators on the two FPGAs are shown in Table \ref{tab:config-and-utilization}. We use the Xilinx Vitis 2020.2 for hardware synthesis and place \& route. The reported resource utilization and frequency are obtained after place \& route.

\subsection{Evaluation Results}

\noindent \textbf{Effect of Model Optimization}: To compare the performance of our proposed model-architecture co-design, we first evaluate the effect of our simplified models on a single CPU core. Table \ref{tab:compacc} shows the reduction in the number of operations and the loss in Average Precision (AP) of the simplified models. Our neighbor pruning strategy achieves near-linear reduction in the memory accesses and computation with respect to the number of neighbors left. The simplified attention mechanism greatly reduces half of the total computation while the LUT-based time encoder reduces another 15\%. On a single CPU core, our simplified model achieves an average of $3.21\times$ speedup in throughput with less than 0.0033 drop in AP. Note that our LUT-based time encoder does not show evident improvement due to the hardware limitation that CPU cannot store the LUT on-chip for fast access.

\noindent \textbf{Cross-Platform Comparison}: We compare the performance of our FPGA accelerators with the same optimized code as in Section \ref{sec:case} on CPU (32 threads) and GPU. 
On the FPGA platforms, we run three models with different sizes NP(L/M/S).
Due to the limited size of external memory, we only test the performance of the Wikipedia dataset on Xilinx ZCU104.
Figure \ref{fig:main} shows the latency and throughput of the baselines and our hardware accelerators. 
We show the latency and throughput under various batch sizes (the first two columns).
Our accelerator on Xilinx Alveo U200 achieves more than $13.9/15.8/17.9\times$ speedup compared with CPU and more than $4.6/5.2/6.0\times$ speedup compared with GPU using the NP(L/M/S) models. 
Note that the accuracy of our simplified models are the same on FPGAs as on CPU (see Table \ref{tab:compacc}).
The obtained speedup on FPGAs is due to our model-architecture co-design that results in low complexity and takes advantages of the FPGA features (Section \ref{sec:Model-Architecture Co-Design}).
To simulate the performance when deployed in production environments, we also evaluate the real-time latency of inferencing on the upcoming graph signals every 15 minutes on the three datasets. We perform inference on batches of new edges in every 15 minutes and measure the latency of each batch.
On Xilinx ZCU104, our hardware accelerators achieve similar latency as GPU but with larger fluctuation due to the limited computing resources.
On Xilinx Alveo U200, we achieve remarkable speedup (more than $3.66\times$) in latency compared with GPU. Our smallest model NP(S) has less than 10ms latency on all three datasets which meets the requirements of most real-time applications.

\begin{figure}[t]
    \centering
    \input{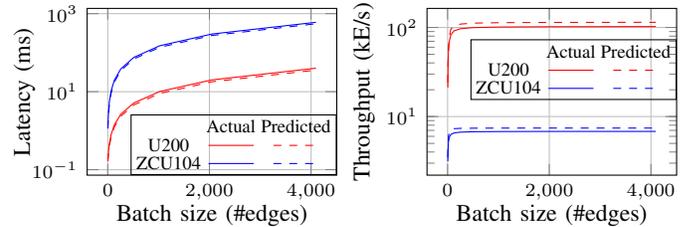}
    \caption{Predicted and actual performance on two FPGAs with the NP(M) model on the Wikipedia dataset.}
    \label{fig:perfmodel}
\end{figure}
\noindent \textbf{Evaluation of Performance Model}: We evaluate our performance model by comparing the predicted latency and throughput with the experimental results.
On average, the prediction error ranges from $9.9\%$ to $12.8\%$. 
Figure \ref{fig:perfmodel} shows the prediction error using our performance model on the Wikipedia dataset. 
We attribute the prediction error to two reasons: (1) the fine-grained hardware pipelines generated by the Xilinx Vitis have some flushing \& draining cycles that are not included in the performance model. This is hard to estimate since the extra cycles are usually decided by the platforms and the version of the compiler. (2) The refreshing behavior of the DDR memory is hard to predict which leads to periodic extra memory latency.  

\begin{figure}[t]
    \centering
    \input{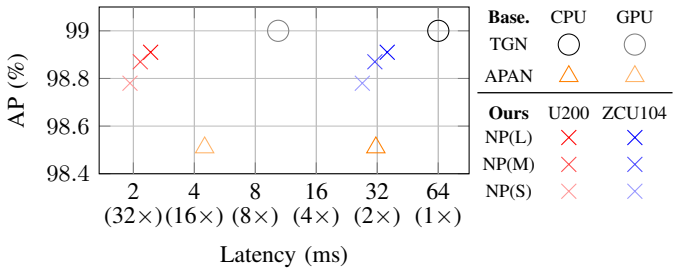}
    \caption{Accuracy and latency on the Wikipedia dataset (batch size 200).}
    \label{fig:acc-lat}
\end{figure}

\noindent \textbf{Comparison with State-of-the-Art Method}: Figure \ref{fig:acc-lat} shows the accuracy-latency curve of our model-architecture co-design, our baseline method TGN \cite{tgn}, and latency-targeted TGNN APAN \cite{apan}. We show the accuracy and latency of TGN and APAN on CPU and GPU since there is no dedicated hardware accelerator designed for these methods. Our co-design achieves significantly higher accuracy than APAN while maintaining similar latency using ZCU104 with CPU and around $2\times$ improvement in latency using U200 with GPU.

\section{Related Works}

The acceleration of GNN on static graphs has been extensively studied before. For example, researchers have proposed algorithm optimizations such as pruning \cite{channelpruning,Xu2020Dynamically} and compression \cite{Wang_2021_CVPR} techniques to reduce the computation complexity. However, the only work on accelerating TGNNs that we are aware of is APAN \cite{apan} which reduces the latency by asynchronous processing. 
On the other hand, there are many hardware accelerators proposed for GNNs on static graphs. HyGCN proposed \cite{yan2020hygcn} hybrid accelerators for GNN with a self-adaptive sliding window technique to reduce the memory traffic. GraphACT \cite{graphact} proposed an FPGA accelerator for sub-graph sampling-based GNN training. AWB-GCN \cite{geng2020awb} proposed a dynamic re-balancing technique to resolve the workload in-balance in GNN inference. Nevertheless, the previous works exploit massive data parallelism and task parallelism on static graphs which are not designed to process the vertex memory updating or deal with temporal dependency on dynamic graphs. To the best of our knowledge, this is the first work to optimize the TGNN inference through comprehensive model-architecture co-design.

\section{Conclusion}

In this work, we proposed a model-architecture co-design for temporal GNN on FPGA platforms.
We defined performance metrics and conducted a case study to identify the bottlenecks in the existing TGNN inference methods. 
We designed a light-weight and hardware-friendly TGNN inference algorithm which has low computation complexity and external memory accesses which takes advantage of the programmable on-chip memory and massive data parallelism of FPGAs.
We mapped the optimized models to principled hardware architectures and implemented the corresponding hardware accelerators on two FPGAs. 
Our co-design on FPGA platforms achieved significantly better performance than state-of-the-art methods on CPU and GPU on real-world datasets.

\bibliographystyle{IEEEtran}
\bibliography{main_short}

\begin{thebibliography}{10}
\providecommand{\url}[1]{#1}
\csname url@samestyle\endcsname
\providecommand{\newblock}{\relax}
\providecommand{\bibinfo}[2]{#2}
\providecommand{\BIBentrySTDinterwordspacing}{\spaceskip=0pt\relax}
\providecommand{\BIBentryALTinterwordstretchfactor}{4}
\providecommand{\BIBentryALTinterwordspacing}{\spaceskip=\fontdimen2\font plus
\BIBentryALTinterwordstretchfactor\fontdimen3\font minus
  \fontdimen4\font\relax}
\providecommand{\BIBforeignlanguage}[2]{{%
\expandafter\ifx\csname l@#1\endcsname\relax
\typeout{** WARNING: IEEEtran.bst: No hyphenation pattern has been}%
\typeout{** loaded for the language `#1'. Using the pattern for}%
\typeout{** the default language instead.}%
\else
\language=\csname l@#1\endcsname
\fi
#2}}
\providecommand{\BIBdecl}{\relax}
\BIBdecl

\bibitem{channelpruning}
H.~Zhou, A.~Srivastava, H.~Zeng, R.~Kannan, and V.~Prasanna, ``Accelerating
  large scale real-time gnn inference using channel pruning,'' \emph{Proc. VLDB
  Endow.}, 2021.

\bibitem{Xu2020Dynamically}
X.~Xu, W.~Feng, Y.~Jiang, X.~Xie, Z.~Sun, and Z.-H. Deng, ``Dynamically pruned
  message passing networks for large-scale knowledge graph reasoning,'' in
  \emph{ICLR}, 2020.

\bibitem{Wang_2021_CVPR}
J.~Wang, Y.~Wang, Z.~Yang, L.~Yang, and Y.~Guo, ``Bi-gcn: Binary graph
  convolutional network,'' in \emph{IEEE/CVF Conference on Computer Vision and
  Pattern Recognition (CVPR)}, 2021.

\bibitem{graphact}
H.~Zeng and V.~Prasanna, ``Graphact: Accelerating gcn training on cpu-fpga
  heterogeneous platforms,'' in \emph{Proceedings of the 2020 ACM/SIGDA
  International Symposium on Field-Programmable Gate Arrays}, 2020.

\bibitem{yan2020hygcn}
M.~Yan, L.~Deng, X.~Hu, L.~Liang, Y.~Feng, X.~Ye, Z.~Zhang, D.~Fan, and Y.~Xie,
  ``Hygcn: A gcn accelerator with hybrid architecture,'' in \emph{IEEE
  International Symposium on High Performance Computer Architecture (HPCA)},
  2020.

\bibitem{geng2020awb}
T.~Geng, A.~Li, R.~Shi, C.~Wu, T.~Wang, Y.~Li, P.~Haghi, A.~Tumeo, S.~Che,
  S.~Reinhardt \emph{et~al.}, ``Awb-gcn: A graph convolutional network
  accelerator with runtime workload rebalancing,'' in \emph{IEEE/ACM
  International Symposium on Microarchitecture (MICRO)}, 2020.

\bibitem{ying2018graph}
R.~Ying, R.~He, K.~Chen, P.~Eksombatchai, W.~L. Hamilton, and J.~Leskovec,
  ``Graph convolutional neural networks for web-scale recommender systems,'' in
  \emph{ACM SIGKDD International Conference on Knowledge Discovery Data Mining
  (KDD)}, 2018.

\bibitem{zhang2019star}
J.~Zhang, X.~Shi, S.~Zhao, and I.~King, ``Star-gcn: Stacked and reconstructed
  graph convolutional networks for recommender systems,'' in \emph{IJCAI},
  2019.

\bibitem{jin2020recurrent}
W.~Jin, M.~Qu, X.~Jin, and X.~Ren, ``Recurrent event network: Autoregressive
  structure inference over temporal knowledge graphs,'' in \emph{Conference on
  Empirical Methods in Natural Language Processing (EMNLP)}, 2020.

\bibitem{sedyt}
H.~Zhou, J.~Orme-Rogers, R.~Kannan, and V.~Prasanna, ``Sedyt: A general
  framework for multi-step event forecasting via sequence modeling on dynamic
  entity embeddings,'' in \emph{ACM International Conference on Information and
  Knowledge Management (CIKM)}, 2021.

\bibitem{xianyu}
A.~Li, Z.~Qin, R.~Liu, Y.~Yang, and D.~Li, ``Spam review detection with graph
  convolutional networks,'' \emph{ACM International Conference on Information
  and Knowledge Management (CIKM)}, 2019.

\bibitem{tgat}
da~Xu, chuanwei ruan, evren korpeoglu, sushant kumar, and kannan achan,
  ``Inductive representation learning on temporal graphs,'' in
  \emph{International Conference on Learning Representations}, 2020.

\bibitem{tgn}
E.~Rossi, B.~Chamberlain, F.~Frasca, D.~Eynard, F.~Monti, and M.~Bronstein,
  ``Temporal graph networks for deep learning on dynamic graphs,'' in
  \emph{ICML 2020 Workshop on Graph Representation Learning}, 2020.

\bibitem{dysat}
A.~Sankar, Y.~Wu, L.~Gou, W.~Zhang, and H.~Yang, ``Dysat: Deep neural
  representation learning on dynamic graphs via self-attention networks,'' in
  \emph{International Conference on Web Search and Data Mining}, 2020.

\bibitem{evolvegcn}
A.~Pareja, G.~Domeniconi, J.~Chen, T.~Ma, T.~Suzumura, H.~Kanezashi, T.~Kaler,
  T.~B. Schardl, and C.~E. Leiserson, ``{EvolveGCN}: Evolving graph
  convolutional networks for dynamic graphs,'' in \emph{AAAI Conference on
  Artificial Intelligence}, 2020.

\bibitem{dyrep}
R.~Trivedi, M.~Farajtabar, P.~Biswal, and H.~Zha, ``Dyrep: Learning
  representations over dynamic graphs,'' in \emph{International Conference on
  Learning Representations}, 2019.

\bibitem{apan}
X.~Wang, D.~Lyu, M.~Li, Y.~Xia, Q.~Yang, X.~Wang, X.~Wang, P.~Cui, Y.~Yang,
  B.~Sun \emph{et~al.}, ``Apan: Asynchronous propagation attention network for
  real-time temporal graph embedding,'' in \emph{Proceedings of the 2021
  International Conference on Management of Data}, 2021, pp. 2628--2638.

\bibitem{wang2021inductive}
Y.~Wang, Y.-Y. Chang, Y.~Liu, J.~Leskovec, and P.~Li, ``Inductive
  representation learning in temporal networks via causal anonymous walks,'' in
  \emph{International Conference on Learning Representations}, 2021.

\bibitem{transformer}
A.~Vaswani, N.~Shazeer, N.~Parmar, J.~Uszkoreit, L.~Jones, A.~N. Gomez,
  L.~Kaiser, and I.~Polosukhin, ``Attention is all you need,'' in
  \emph{International Conference on Neural Information Processing Systems},
  2017.

\bibitem{hinton2015distilling}
G.~Hinton, O.~Vinyals, and J.~Dean, ``Distilling the knowledge in a neural
  network,'' \emph{stat}, p.~9, 2015.

\bibitem{lu2021demystifying}
A.~Lu, Z.~Fang, W.~Liu, and L.~Shannon, ``Demystifying the memory system of
  modern datacenter fpgas for software programmers through microbenchmarking,''
  in \emph{ACM/SIGDA International Symposium on Field-Programmable Gate
  Arrays}, 2021.

\end{thebibliography}

\end{document}